\documentclass[epsf]{aa}
\usepackage{graphics}
\begin{document}
\def\cd{cd$^{-1}$}
\def\cds{cd$^{-1}$\,}
\def\kms{km~s$^{-1}$}
\def\kmss{km~s$^{-1}$\,}
\title{Simultaneous intensive photometry and high resolution spectroscopy
of $\delta$ Scuti stars.\thanks{Based on
observations collected at ESO-La Silla (Proposals 57.E-0162, 61.E-0120)}}
\subtitle{V. The high--degree modes in the pulsational content
of BV Circini}
\author{L. Mantegazza, E. Poretti \and F.M. Zerbi}
\offprints{L. Mantegazza}
\institute{Osservatorio Astronomico di Brera, Via E.~Bianchi 46,
       I--23807 Merate, Italy \\
luciano$\backslash$poretti$\backslash$zerbi@merate.mi.astro.it}
\date{Received date; accepted date }

\abstract{
We discuss here the pulsation properties of the $\delta$ Scuti star BV Circini
on the basis of data obtained during a simultaneous photometric and spectroscopic
campaign in 1996 and a spectroscopic one in 1998, and taking also advantage
of the previous photometric observations by Kurtz (1981).
Nine pulsation modes were detected from photometry and 
thirteen from spectroscopy; five of them are in common to both techniques.
The spectroscopic data give ample evidence of dramatic amplitude
variations in some modes, in particular the strongest spectroscopic mode in
1998 was not detectable in 1996 data. The two dominant photometric modes
(6.33 and 7.89 \cd) are observed on both seasons. 
The typing of the modes was performed by means of a simultaneous model
fit of line profile and light variations. The 6.33 \cds photometric term 
is probably the fundamental radial mode, while the 7.89 \cds is a 
nonradial mode with $m\neq$0. 
There are six
high--degree prograde modes with an azimuthal order $m$ ranging from --12 to --14,
and also a retrograde mode with $m\sim 7$. These modes combined with the
identification of the 6.33 \cds mode allowed us to estimate $i\sim 60^o$
for the value of 
the inclination of the rotation axis. An accurate evaluation of the main
stellar physical parameters is also proposed as a result of the pulsational analysis.
}
\authorrunning{Mantegazza et al.}
\titlerunning{BV Cir}
\maketitle

\section{Introduction}

Asteroseismology is knowing an increasing interest thanks to
satellites which are expected to fly in the incoming years
(MOST, MONS, COROT) and which will perform accurate investigations of
stellar variability at the $\mu$mag level. Stars located in the lower
part of the instability
strip are suitable candidates to find a large quantity of excited modes,
both radial and nonradial. The detailed knowledge of the pulsational properties
of $\delta$ Sct stars can help in the preparation of the scientific
background of these missions; a review of the photometric properties
of these variable stars is given by Poretti (2000). The main difficulty
in their investigation is the typing of the excited modes, i.e. classifying
the oscillations in terms of quantum numbers $n, \ell$ and $m$. To proceed
on this way, we started an observational project combining photometric and
spectroscopic techniques at the European Southern Observatory.
In this paper we supply the last result, after those
on FG Vir (Mantegazza et al. 1994), X Cae (Mantegazza \& Poretti 1996,
Mantegazza et al. 2000) 
and HD~2724 (Mantegazza \& Poretti 1999). 

BV Circini$\equiv$ HD 132209 was discovered to be a $\delta$ Scuti
variable star by Kurtz (1981). He detected four pulsation modes from the
frequency analysis of 18 nights of observations in the
$B$ and $V$ colours in 1980. The author also pointed out that these modes did not
provide a complete description of the stellar variability. The strongest term
at 6.328 \cd\, was suggested to be the radial fundamental mode.
Since the star is rather bright ($V=6.56$) and its light variations are quite
large ($\Delta V>0.1$ mag), we decided to include it among our targets.
\section{Physical parameters}

The stellar physical parameters can be estimated independently by means
of the $uvby\beta$ and Geneva colour indices and related calibrations.
They are required for fitting
the line profile variations and are useful for discussing the pulsation
properties.

The Moon \& Dworetsky (1985) calibration applied to Hauck \& Mermilliod's
$uvby\beta$
data supplies $T_{\rm eff}=7330\pm30$~K and $\log g=3.58\pm 0.06$
(this estimate also includes the correction for metallicity effects as derived
 by Dworetsky \& Moon 1986). The recent calibration of Geneva photometric indices
by Kunzli et al. (1997) applied to the data by Rufener (1988)
supplies $T_{\rm eff}=7185\pm62$~K and $\log g=3.95\pm 0.10$. The
discrepancy in the value of the temperature is negligible since models
are marginally affected by small temperature differences. In the
following an average value of $T_{\rm eff}=7260\pm100$~K will be
adopted. 

The disagreement on $\log g$ deserves further investigation in the future,
since it has also 
been found for X Cae (Mantegazza et al. 2000). From some tests
made on other $\delta$ Sct stars, the $\log g$ values derived from Geneva
photometry seem to be always higher than those derived from $uvby\beta$
photometry,
even if usually with a lesser extent than here. In the case of BV Cir,
the {\sc hipparcos} parallax measurement ($7.85\pm0.60$ mas) offers us
the possibility to solve the matter. From this
parallax 
and the interstellar reddening estimate supplied by the $uvby\beta$ indices
($E(b-y)=0.022$, and thus $A_V=0.095$), we derive $M_V=0.94\pm0.17$ and
therefore $M_{\rm bol}=0.97$ ($BC=+0.035$, Flower 1996).
Finally, from the photometric temperature and the {\sc hipparcos} luminosity,
we get $R=3.59\pm 0.30$ R$_{\sun}$.
BV Cir is then an evolved $\delta$ Scuti star, located 
in the middle of the instability strip, slightly shifted 
toward the cold border.

 According to the theoretical models by
Schaller et al. (1992) a mass
of $2.0\pm0.1$ M$_{\sun}$ can correspond to such temperature and luminosity;
finally, by combining this with the radius estimate we get
$\log g=3.62\pm 0.09$ in nice agreement with the estimate from
the Moon \& Dworetsky (1985) calibration.
\section{Frequency analysis of Kurtz's (1981) photometric data}

Since the solution proposed by Kurtz (1981) did not
give not a fully satisfactory fit of the observations, we decided to analyze
his data with the least--squares power spectrum technique, which has been
widely used by us in a lot of papers and has proved to be rather efficient
in analysing multi--periodic time--series (e.g. Mantegazza et al. 2000).

Both $B$ and $V$ datasets were independently analyzed and the
results are summarized in Tab.~\ref{frk}. Nine components were found in
common between the two datasets, 8 with the same frequencies and
one with an uncertainty of 1 \cds (9.232 \cds in $B$ and 10.232
\cds in $V$). One low frequency term was detected in both datasets,
but the frequency is slightly different (0.295 \cds in $B$ and
0.239 \cds in $V$). Owing to the fact that the data are
distributed in 5 sequential segments and were obtained with three
different telescopes, this term could be due to the procedure
adopted to align the differents datasets.
This spurious nature can account for the quite high $A_B/A_V$ ratio,
i.e. 1.9, unlikely for a pulsation mode.

The four strongest terms are the same as detected by Kurtz (1981). We
note that among the new terms,  a frequency (9.901 \cd) is very
close to a 2 \cds alias of the term at 7.892 \cd. However, it
should be a real feature since each attempt to get rid of it has
failed (for example by substituting the two frequencies with the
intermediate alias at 8.90 \cds and so on).

Other terms are probably present but their unambiguous detection
is problematic with the present data. There could be also variations in the
amplitudes of the detected terms, as detected in the spectroscopic data (see below);
since the photometric baseline spans 116 days, these variations could be
appreciable, but again the data sample is inadequate to supply
an unambiguous answer.

In order to get a final list of frequencies, the $B$ and $V$ data were
put together by aligning the zero--points and rescaling the $V$ data
in order to match the $B$ amplitudes; the resulting time-series
was frequency analyzed again. This approach is the same as used by
Breger et al. (1998) to study FG Vir and by Mantegazza et al.
(2000) to study X Cae. The frequencies detected in this way
are also listed in
Tab.~\ref{frk}; the $B$ and $V$ amplitudes obtained by fitting the
respective time--series and the rms residual are also reported.
\begin{table}
\caption[]{Reanalysis of the photometric  data supplied by Kurtz (1981).
The terms at 11.077 and 11.128 \cds were no further observed in 1996.}
\begin{flushleft}
\begin{tabular}{rrrrrr}
\hline
\multicolumn{3}{c}{Frequencies [\cd]}&&\multicolumn{2}{c}{Amplitudes [mmag]}\\
\cline{1-3} \cline{5-6}
$B$ & $V$ & $B\&V$ & & $A_B$ & $A_V$ \\
\noalign{\smallskip}
\hline
0.295 & 0.239 & 0.294& & 3.1& 1.6\\
6.328 & 6.328 & 6.328 &&24.3& 19.7\\
7.892 & 7.892 & 7.892& &10.0& 8.2\\
9.901 & 9.901 & 9.901& & 5.6& 4.6\\
9.232 & 10.233 & 10.234& & 2.5&2.0\\
11.077 & 11.077 & 11.077& &3.5&3.6\\
11.128 & 11.128 & 11.128 &&6.8&6.0\\
11.631 & 11.631 & 11.631 &&7.3&6.0\\
12.289 & 12.289 & 12.289 &&12.5&10.0\\
12.381 & 12.380 & 12.381 &&5.4&4.1\\
\hline
\multicolumn{3}{c}{rms residuals}&&4.4&3.5\\
\hline
\end{tabular}
\end{flushleft}
\label{frk}
\end{table}

\section{New photometric observations}

The photometric data presented in this paper were collected during
a campaign in June 1996 with the 0.5~m telescope at la Silla
Observatory. During five consecutive nights 
322 and 312 measurements were collected in the Str\"omgren $v$ and
$y$ bands respectively. 
Differential photometry was performed by using HD~132249 and HD~132222
as comparison stars.  These stars were used by Kurtz (1981), who found
them constant at the mmag level.
Due to the non-perfect atmospheric
transparency during these Chilean winter nights, systematic
intra-night variations were noticed in the extinction coefficient.
Therefore, the data were reduced by means of the instantaneous extinction
coefficient procedure (Poretti \& Zerbi 1993). Nevertheless,
the photometric measurements could not reach the typical level of
precision for the La Silla site; the magnitude differences between
the two comparison stars show a standard deviation of 6.5 mmag in
$y$--light and 8.7 mmag in $v$--light.

The new observations were analyzed with the same approach adopted
for the Kurtz (1981) data: the $v$ and $y$ measurements were first
frequency analyzed separately and then were merged together. The
detected frequencies are listed in the first 3 columns of Tab.~\ref{fru}.
We notice that the spectral resolution of this dataset, 0.2 \cd,
is consistently lower than that of the 1980 dataset. In the 4th 
column the corresponding frequencies of
Tab.~\ref{frk} are reported for comparison purposes. The frequencies at 6.328
and 7.892 \cds that were dominant in the 1980 data are still present
in the new photometry still with a dominant character. A number of
other terms such as 9.901, 11.631, 12.289, 12.381 \cds were also
confirmed in the new photometry while the term at 10.234 \cds
presented some aliasing uncertainty. The terms at 11.077 and
11.128 \cds were not recovered in the 1996 dataset;
as we shall see later, they were not  detected 
in the spectroscopic analysis too.
Also in our dataset low frequency terms are observed, at
values (0.36 \cds and its alias 0.64 \cd)
different from those observed in Kurtz's data (0.29 \cd): this
difference can be accounted for the worst frequency resolution, but
it is greater than that observed for other terms (see Tab.~\ref{fru}). Moreover,
the power spectrum of the measurements between the two comparison
stars shows increased noise at $f<$1~\cd, strongly supporting a spurious nature
of the detected peaks.

\begin{figure}
\resizebox{\hsize}{!}{\includegraphics{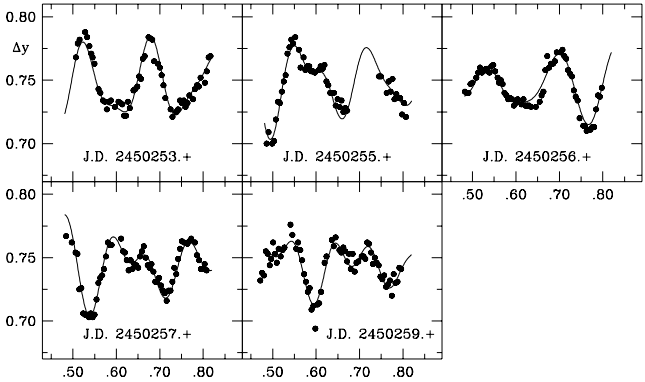}}
\caption[ ]{Light curves obtained at ESO in the $y$--light. The fit
is calculated by using the parameters listed in Tab.~2}
\label{lc}
\end{figure}

Amplitudes in the $v$ and $y$ bands were derived by fitting the
respective data adopting the Kurtz values for the frequencies,
which are more accurate because of the longer baseline. Figure~\ref{lc} 
shows the fit of the measurements in the $y$--band. By comparing the $V$
amplitudes of Kurtz's (1981) data with those of the present ones
it appears that most of the terms have comparable values. This
is not true for the 10.23 \cds term, much stronger in the recent
data. However,  this term is not well resolved from the alias at
2~\cds of the 12.29 \cds term and therefore this effect could be
spurious.

\begin{table}
\caption[]{Frequency analysis of the new photometric data.}
\begin{flushleft}
\begin{tabular}{rrrrrrr}
\hline
\multicolumn{4}{c}{Frequencies [\cd]}&&\multicolumn{2}{c}{Amplitudes [mmag]}\\
\cline{1-4} \cline{6-7}
$v$ & $y$ & $v\&y$&$B\&V$ &&  $A_v$ & $A_y$ \\
\noalign{\smallskip}
\hline
0.64 & 0.36 & 0.65 & ---  & & 4.8  & 3.9\\
6.33 & 6.33 & 6.33 & 6.328 && 28.2 & 19.3\\
7.88 & 7.88 & 7.87 & 7.892 && 16.0 & 10.3\\
---  & 9.93 & 9.94 & 9.901 && 6.9  & 5.5\\
10.28& 9.23 & 9.27 & 10.234&& 9.2  & 7.1\\
11.69&11.61 &11.63 & 11.631&& 8.7  & 5.8\\
---  &11.25 & 12.29& 12.289&& 10.5 & 7.3\\
11.38&13.42 & 12.36& 12.381&& 7.3  & 5.4\\
\hline
&\multicolumn{4}{c}{rms residuals}& 7.2  & 5.4\\ \hline
\end{tabular}
\end{flushleft}
\label{fru}
\end{table}

\section{Spectroscopic observations and data processing}

The spectroscopic observations were performed during two runs in 1996
and 1998 at La Silla Observatory (ESO) with the Coud\'e Echelle Spectrograph
 attached
to the Coud\'e Auxiliary Telescope (1.4~m).

The first run  (6 consecutive nights, June 18--24,
1996) was performed in Remote Control Mode from Garching with the CES
configured in the blue path with the long camera and the CCD \#38. The
reciprocal dispersion was 0.018~\AA~pix$^{-1}$ with an effective resolution
of about 54000, and the useful spectral region ranged from 4482 to 4532 \AA.
The weather conditions were not always optimal and, to get
an adequate $S/N$, the integration times ranged between 12 to 25 minutes.
118 useful spectrograms were gathered; they monitor the star
variability for about 38 hours (duty cycle of about 30$\%$) and on a
baseline of 5.2 days.

The second run (12 consecutive nights, June 7--19, 1998) was performed from
La Silla with the same CCD and blue path, but with the very--long camera.
The reciprocal dispersion was of 0.0075~\AA~pix$^{-1}$ with an
effective resolution of about 51000. The useful spectral region ranged from
4498 to 4517 \AA. Integration times were of 15 minutes and a total of 156
useful spectrograms were obtained during 9 nights, corresponding to about
48 hours of stellar monitoring (duty cycle about 18$\%$) on a baseline
of 11.3 days.

The data reduction was performed using the {\sc midas} package developed at ESO.
The spectrograms have been normalized by means of internal quartz lamp flat
fields,
and calibrated into wavelengths by means of a thorium lamp.

The spectrograms of each run have then been averaged and some windows on
the continuum were selected in the very high $S/N$ mean spectra. The
individual spectrograms have been normalized to the continuum by
fitting a 3$^{\rm rd}$ degree polynomial to these windows for the first run set
(longer spectral range) and a 2$^{\rm nd}$ degree polynomial for the second run set
(shorter spectral range).

Finally the spectra have been shifted in order to remove the
observer's motion (Earth's rotation and revolution). 
The spectrograms were rebinned with a step of 0.04 \AA~ (average of 2
original pixels) for the first season data and of 0.035 \AA~ 
(average of 5 original pixels) for the second one. This step allows us to save
the effective resolution according to the Nyquist criterion, and not to have
too many pixels to analyze.

Due to the relatively high projected rotation velocity (see below), only the
Fe~{\sc ii} line at 4508.3 \AA~ is completely free from blends of adjacent
features, and therefore was the only line suitable to study line profile variations.

The mean standard deviation of the pixels belonging to the continuum allows
an estimation of the $S/N$ of the spectrograms. The resulting average value
at the continuum level is about 257 for the first season data and 251 for the
second one.

A non--linear least--squares fit of a rotationally broadened Gaussian profile
to the average line profiles of the two seasons allows us to get very similar
estimates of the projected rotational velocity and the intrinsic width
of the 4508~\AA~ line. We get $v\sin i=96.5\pm 1.0$ \kms and $W_i=12.5\pm 0.5$
\kms, respectively.

\section{Analysis of spectroscopic variations}

\subsection{Radial velocities and the second moment of the line profile}

Radial velocities were derived by the computation of the instantaneous
line barycentre (i.e. they coincide with the first moment of the line profile,
 Balona 1986). They were derived for both datesets and the two
resulting timeseries were analyzed by the same least--squares
technique adopted for the light curve analysis.
The amplitudes of radial velocity variations are rather small, 
not exceeding 5 \kmss peak to peak; 
the accuracy of these data is 0.6--0.7 \kmss as estimated from the white 
noise level. Because of this, only the 6.33 \cds can be detected unambiguously 
in both datasets. Its semi-amplitude is the same within the uncertainties 
in both seasons: $1.3\pm0.1~$\kms. The fit of the radial velocity data with 
the other 
photometrically dominant frequencies shows that these terms should have 
amplitudes of the order of 0.5 \kmss or less.
The analysis of the second moment supplies again the same result for both 
seasons: only one mode is clearly discernible, but it  
is the 7.89 \cd, not the 6.33 \cds mode. Forcing the fit of the 6.33 \cds
 term to the data 
we get a negligible amplitude. The conclusion is that the 6.33 \cd mode, 
which is also the photometrically dominant term, is axisymmetric ($m=0$); 
on the contrary the 7.89 \cds mode should have $m\neq 0$.

\subsection {Analysis of the line profile variations}

The variations of the shape of the line profiles was studied by means
of the pixel by pixel analysis. A detailed description of this approach
was made by Mantegazza \& Poretti (1999) and by Mantegazza (2000).
In particular Fig. 6 of the second paper shows the pixel by pixel
least--squares power spectrum relative to the Fe~{\sc ii} 4508~\AA~ line of BV Cir as derived
from the 1998 data, where it is apparent that there is a significant
 contribution to
the line profile variations in the whole frequency range (from 1 to 20 \cd).
\begin{figure}
\resizebox{\hsize}{!}{\includegraphics{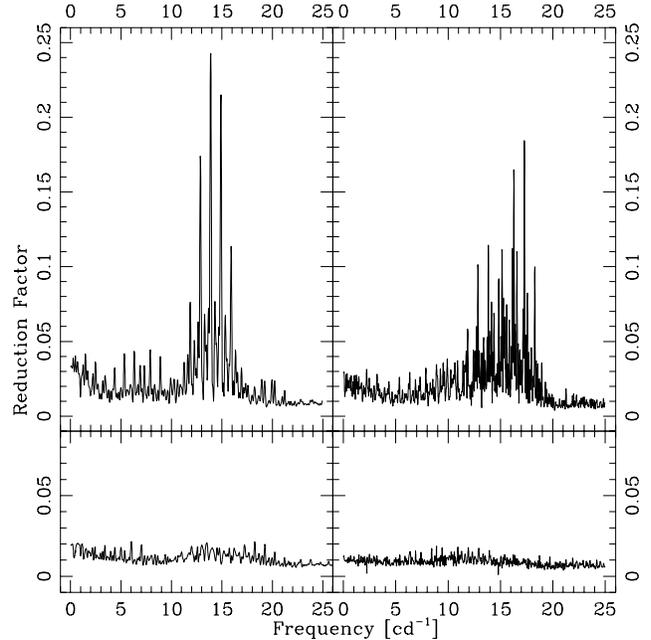}}
\caption[ ]{Least--squares global power spectra of the line profile
variations of the 4508~\AA~ line. Left: 1996 data; right: 1998 data.
Upper panels: spectra of the signal contained in the data. Lower panels:
residual spectra after considering all the detected terms.
The reduction factors  refer to the initial data
variance of the respective season both in the upper and lower panels.}
\label{lsp}
\end{figure}

The frequencies detected for the two datasets with this technique are listed in
order of increasing frequency in Tab.~\ref{frs}.
Not all the terms were detected in both seasons; in particular the dominant
term in the 1998 data (17.28 \cd) was not seen in the 1996 ones.
The upper panels of Fig.~\ref{lsp} show all the signal contained in the data, while
the bottom panels show the signal left after considering all the
detected terms.
In the 2$^{\rm nd}$  and 3$^{\rm rd}$ columns of Tab.~\ref{frs} we report
 the mean square amplitudes
across the line profile of the detected terms for the 1996 and 1998 data
respectively. These quantities are expressed in square units of
the local continuum of the stellar spectrum.
Due to the high number of excited modes and to the complexity of the spectral
window it cannot be ruled out that some of the detected frequencies could be
1~\cds aliases of the true ones. We can be more confident on terms which
have been independently detected in both seasons and/or in the
photometric data.
\begin{table}
\caption[]{Detected modes by pixel by pixel least-squares frequency analysis of
high--resolution spectrograms. The photometric counterparts are also reported.}
\begin{flushleft}
\begin{tabular}{rrrrrrr}
\hline\noalign{\smallskip}
\multicolumn{1}{c}{Freq.}&&\multicolumn{2}{c}{Power}&& \multicolumn{1}{c}{Photom.}\\
\cline{3-4}
\multicolumn{1}{c}{[\cd]} &&1996& 1998&& \multicolumn{1}{c}{[\cd]} \\
\hline
0.65 && 2.88 & 3.28&&0.65  \\
1.48 && 4.02 & 2.88&&---\\
2.22 && ---  & 3.08&&---\\
6.33 && 3.42 & 2.42&&6.328\\
7.89 && 3.40 & 3.55&&7.892\\
9.90 && ---  & 2.85&&9.901\\
11.63&& 3.11 & 3.40&&11.631\\
(13.43)&& 3.18 & ---&&12.381 \\
13.85&& 9.04 & 10.07&&---\\
14.34&& 4.41 & 2.87&&---\\
14.95&& 7.97 & 7.03&&---\\
15.80&& ---  & 3.86&&---\\
16.44&& ---  & 6.74&&---\\
17.28&& ---  & 12.87&&---\\
\hline
\end{tabular}
\end{flushleft}
\label{frs}
\end{table}

Fourteen terms were detected out of which seven were present in both seasons.
Moreover six terms have an independent photometric detection: 0.65, 6.33, 7.89,
9.90, 11.63 and 13.43 \cd. The last value probably corresponds to the 1~\cds alias
of the photometric value of 12.38~\cd and  in the following we shall adopt this
photometric value. The nature of the low--frequency terms is discussed in
Sect.~6.4.

An independent check of these detections can be performed with the
{\sc clean} algorithm (Roberts et al. 1987) generalized to study line
profile variations (see for example Bossi et al. 1998 and De Mey
et al. 1998). Such an approach is complementary to the previous
one because the selection of the peaks is completely automatic,
i.e. free from human judgement, with the consequent merits and
defects. Among the merits the major one is the absence of human
intervention: indeed any time we select a known constituent and
add it to the least--squares solution we intervene on the
representation of the signal thus conditioning to some extent the
next procedure. Among the defects the major one is that
{\sc clean} is in condition to make
a reliable distinction between a true signal and an alias
only when the spectral window is of very good quality.
For this reason the procedure we adopted in the past on similar analyses 
was to put more reliability on the least--squares analysis and use
{\sc clean} only as a cross-check. Moreover, for sake of clarity,
we prefer to use 
the {\sc clean} power spectra to present the frequency content of
the data.

\begin{figure*}
\resizebox{\hsize}{!}{\includegraphics{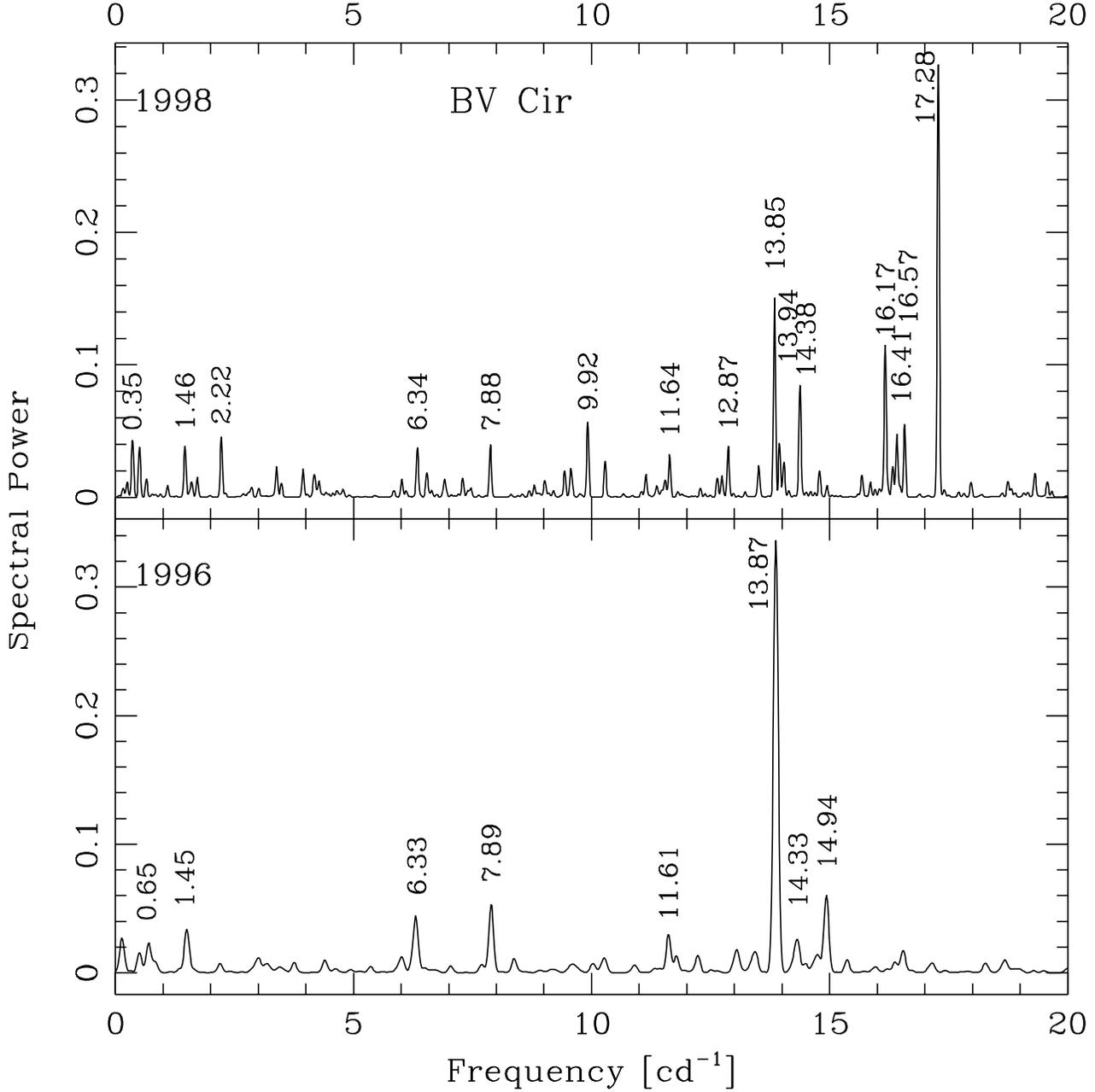}}
\caption[ ]{Average {\sc clean} power spectra of the line profile variations
across the Fe~{\sc ii} 4508~\AA~line. The frequencies of the
main peaks are given in
the labels. Lower panel: 1996 data; upper panel: 1998 data.
}
\label{cps}
\end{figure*}

The results of the {\sc clean} analysis are shown in the two panels of
Fig.~\ref{cps}. By comparing
this figure with Tab.~\ref{frs} we can see that there are no great differences
with the least--squares technique and that both techniques agree on
several detections. Two different alias identifications are observed:
the 0.65 and 0.35 \cds peaks, the 14.94 and 13.94 \cds peaks. Moreover, in
the 1998 data the 12.87 and 13.85 \cds peaks are related to the same
term. 
 It is also possible to see that the peaks of the 1998 data
are narrower than those of the 1996 data because of the longer baseline; the same
fact can be appreciated with the least--squares power spectra (Fig.~\ref{lsp}).

The comparison with the least--squares analysis confirms most of the 
terms (0.65, 1.46, 2.22, 6.33, 7.89, 9.92, 11.63, 14.34 and 17.28 \cd)
and allows us to prefer the 14.94 and 13.85 \cds terms to their aliases.

As regards the terms with frequencies around 16.2--16.5 \cd, both 
{\sc clean} and the least--squares technique confirm the difference between
the 1996 (no significant term detected in this region) and 1998 (a
bunch of detected terms) datasets. The least-squares technique seems
to explain the signal in this region in a simplest way (a unique term,
rather than the three suggested by {\sc clean}). Moreover, {\sc clean} did
not reveal clearly the 15.80 and 12.38 \cds terms.
These small discrepancies can be accounted for by a noise effect and bad
spectral window. In the  following analysis we shall use the terms identified
by means of the least--squares approach.

In particular, we don't put particular
confidence on the low--frequency terms, except for the 1.48 \cds term.
The other terms (0.65 and its alias 0.35 \cd, the 2.22 \cd) are considered
to be spurious ones (see Sect.~6.4).

\begin{figure*}
\resizebox{\hsize}{!}{\includegraphics{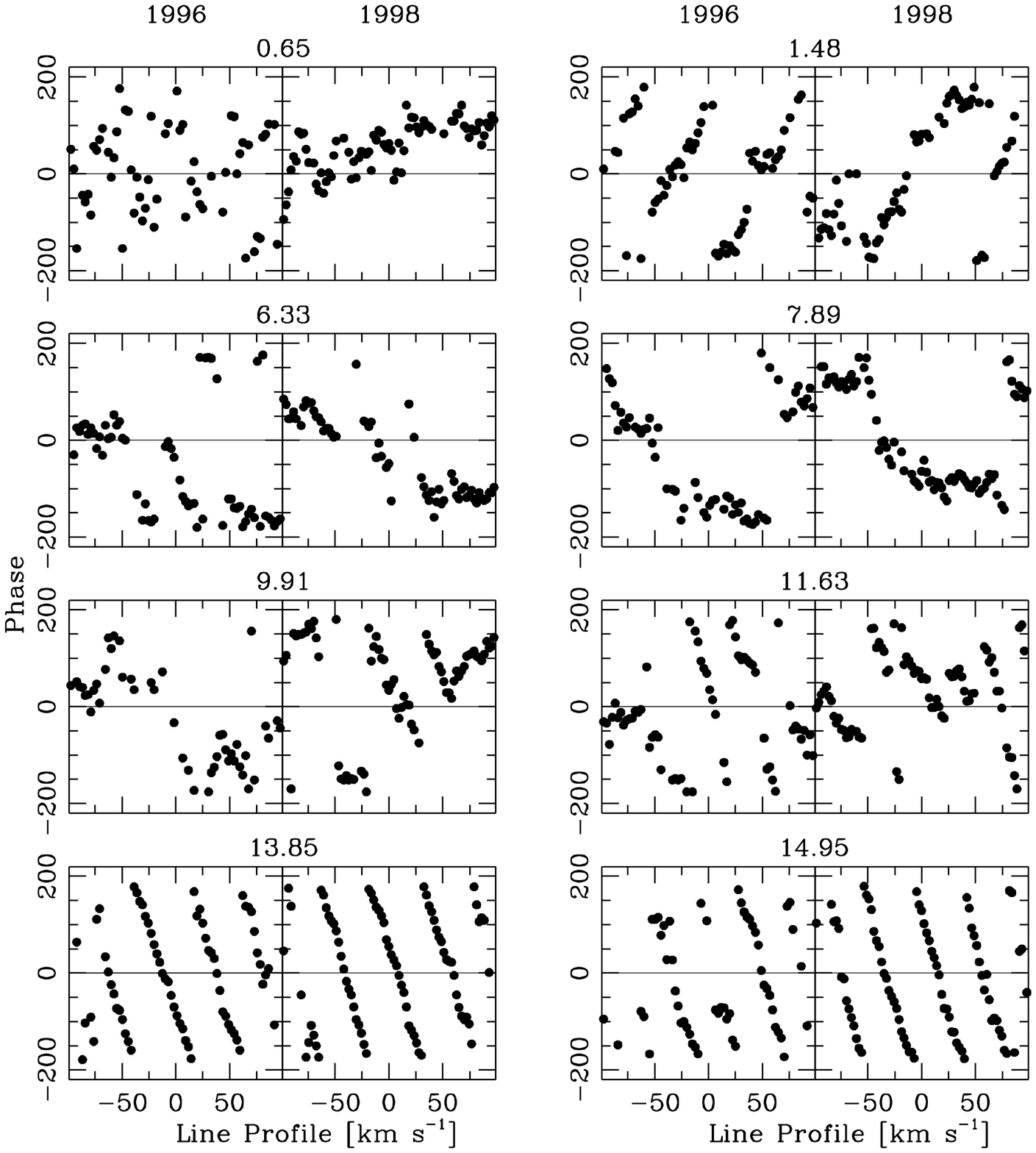}}
\caption[ ]{Behaviours of phases across the line profile
of the Fe~{\sc ii}  4508 \AA~line of the terms detected with
the least--squares power spectrum analysis both in the 1996 and the 1998
 data.  The phases are 
in degrees (each tickmark corresponds to
$50^o$). The thin horizontal line in each panel gives the zeropoint level.
The line profile is measured in Doppler velocities: the centre of the line
defines the zeropoint.
The frequency of each mode is written on the top of each panel. 
The doubtful pulsational nature of the 0.65 \cds term is discussed in the
text.
}
\label{cm}
\end{figure*}

\begin{figure*}
\resizebox{\hsize}{!}{\includegraphics{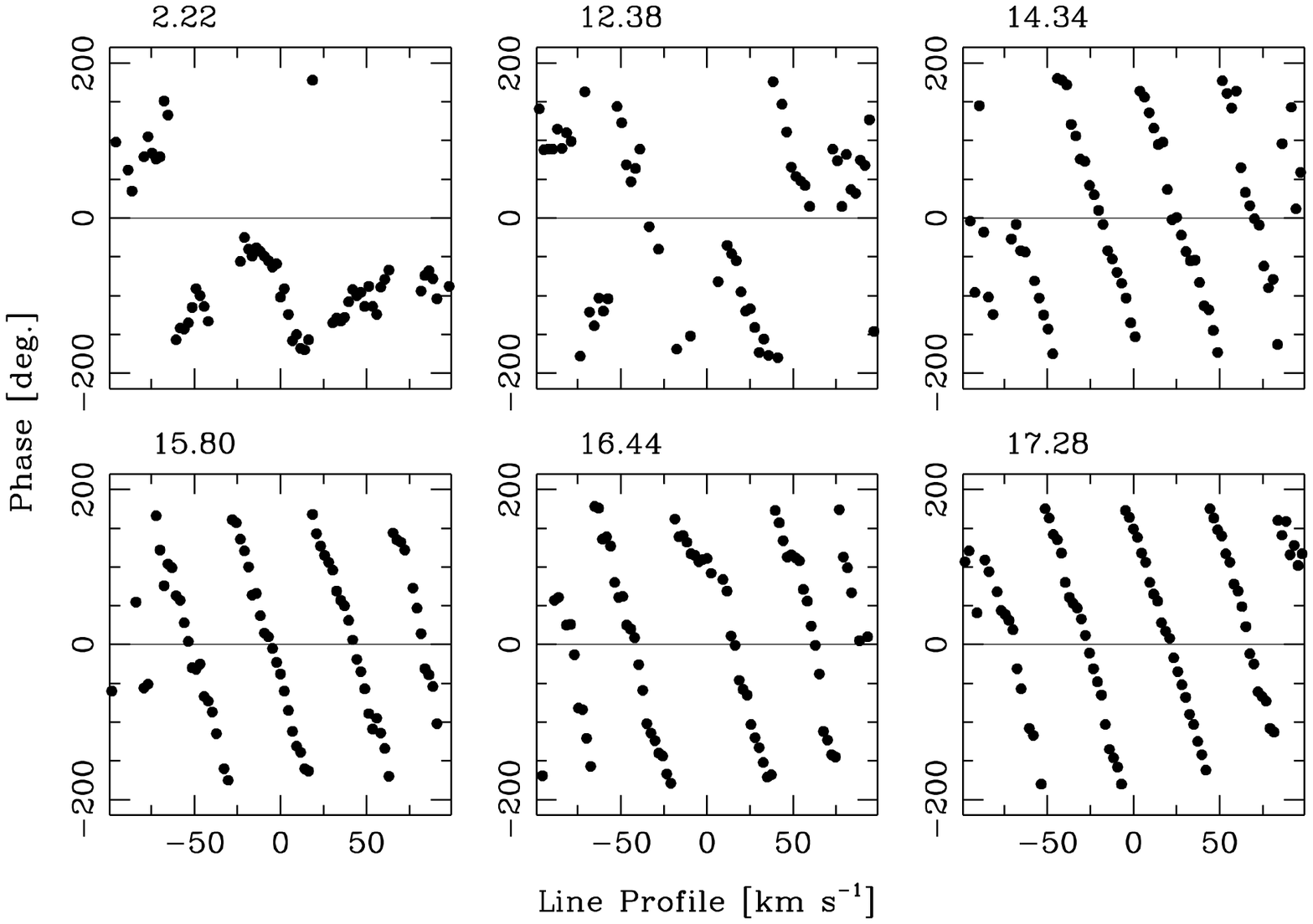}}
\caption[ ]{Same as in the previous figure but relative to the modes
detected in one dataset only: 12.38 \cds and 14.34 \cds were detected in the
1996 data, the others in the 1998 ones. The doubtful pulsational nature of the
2.22 \cds term is discussed in the text.
}
\label{us}
\end{figure*}

The behaviours of phases across the line profile of the modes
adopted for further analysis
are shown in Fig.~\ref{cm} for the 8 modes detected in the spectra
of both seasons and in Fig.~\ref{us} for those detected in one dataset only.
The behaviour of the 9.901 \cds term is puzzling. It was observed
photometrically in 1980 and 1996. On this latter occasion, there was
no significant spectroscopic counterpart (see lower panel of Fig.~\ref{cps}):
the spectroscopic amplitude may be undetectable because the mode
had quite a small photometric amplitude. Moreover, owing to the modest
frequency resolution, its power could have been transferred via the 2~\cds
alias to the stronger 7.89 \cds mode.  The frequency resolution is better
in the 1998 data and the 9.901 \cds term was clearly detected, maybe
thanks to an increased amplitude; unfortunately, we do not have photometric
data with which to check this possibility.  Therefore, we
decided to consider a 9.90 \cds term in the analysis of the
1996 spectroscopic data also.

The behaviour of the phases across the line profile gives already
some hints about the nature of a few of the detected modes.
For instance the phases of the 13.85, 14.34, 14.95, 15.80, 16.44 and 17.28
\cds modes have the typical signature of high--degree prograde modes.
The behaviour of the phases of the 1.48 \cds term is interesting because
it is typical of a retrograde mode.


\subsection{Mode identification: the technique}

It is possible to try to estimate the $\ell$,$m$ parameters of the detected
modes by a model which fits the variations induced on the line profile shape.
The technique adopted to separate the contributions of the different modes
to the global observed line profile variations and to fit them is described
in Mantegazza et al. (2000) and Mantegazza (2000).
The theoretical line profile variations are computed according to the model
developed by Balona (1987), by means of his LNPROF routine. 
If simultaneous light variations are available it is possible to fit them
with the same model at the same time, and in this case a more accurate
typing of the mode can be proposed, since we can constrain at the same time both
the velocity fields and the flux variations.
The best fitting modes are selected according to a ``discriminant''
whose value defines the goodness with which the candidate mode
reproduces the variations of the line profile and of the light (if available)
of the detected frequency term. For some modes the discriminant is also
sensitive to the inclination of the rotational axis, and therefore if
several modes are detected it is possible to build a total discriminant
(the sum of the individual ones) as a function of inclination angle $i$.
The minimum of this function can in principle fix this parameter.

In the case of BV Cir only the 1996 data have simultaneous light
and spectroscopic observations and therefore only these data can be used
to perform a simultaneous fit of both quantities and hence to constrain
velocity and flux variations.

\subsection{Low frequency modes}
Three terms were observed at low frequencies: 0.65, 1.48 and 2.22 \cd.
The attempts to fit 0.65 and 2.22 \cd were fruitless and  no meaningful
identification was possible.
The phase diagrams of 0.65 (Fig.~\ref{cm}, left panel of the upper
row) and 2.22 \cds (Fig.~\ref{us}, left panel of the upper row)
terms look like  a scattered plot, without any discernible trend. This
result supports our hypothesis that these terms are spurious 
effects in the analysis. It can be argued that the 0.65 \cds term is found both in
the spectroscopic and in the photometric datasets, which are independent from
each other. However, there are other peaks in this frequency region  having 
similar amplitudes (see Fig.~\ref{cps}) and therefore the
numeric coincidence can be considered as merely casual. 
We again remind the reader that in the photometric analysis also
there are many doubts on the reality of the 0.65 \cds term.
Therefore, in the discussion,
we prefer not to consider the 0.65 and 2.22 \cds terms as pulsational
modes, even if some uncertainties are left, especially on the former.

\begin{figure}
\resizebox{\hsize}{!}{\includegraphics{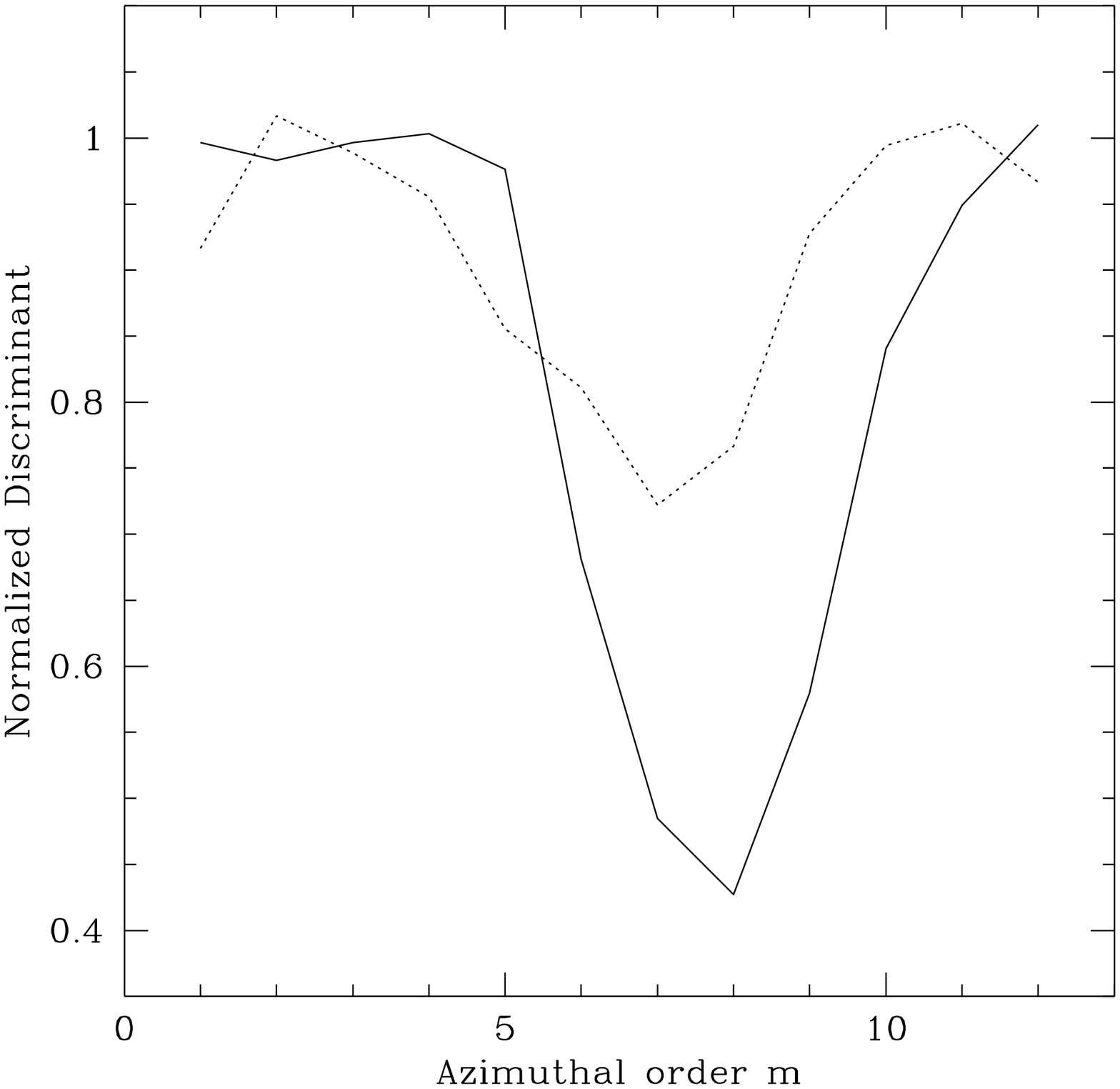}}
\caption[ ]{Discriminants of the 1.48~\cds mode as a function
of the azimuthal order $m$ computed from the 1996 (solid lines) and 1998
(dashed lines) data. These values are relative to $i=55^o$ and $\ell=m$.
But the discriminants are rather insensitive to $i$ and are comparable for
any $\ell\geq m$.}
\label{lfm}
\end{figure}

On the contrary the term at 1.48 \cd, which has been
observed in both seasons, and whose phase diagrams clearly show that it
is a retrograde mode, supplied useful discriminants. As is also evident
from its phase diagrams (Fig.~\ref{cm}, right panel of the upper row) it is
 retrograde and has a rather
high degree, and as in the case of the prograde mode the discriminant is
rather insensitive
to the inclination and to the $\ell$ degree. The derived values from the two
datasets are shown in Fig.~\ref{lfm}.
The 1998 discriminant has a lower depth than the 1996 one because in that
season the $S/N$ was lower. In any case we can estimate $m=$7 or 8.

 Such modes are rarely observed in $\delta$
Scuti stars, and often their detection is rather ambiguous, but in this
case the evidence is remarkable and is independently shown by the two
datasets (Fig.~\ref{cm}). This typing constitutes a warning about the
identification of a low--frequency terms as a $g$--mode, since a $p$--mode
can assume these values in the observer's reference frame.

\subsection{Modes with spectroscopic and photometric
detections}
\begin{figure*}
\resizebox{\hsize}{!}{\includegraphics{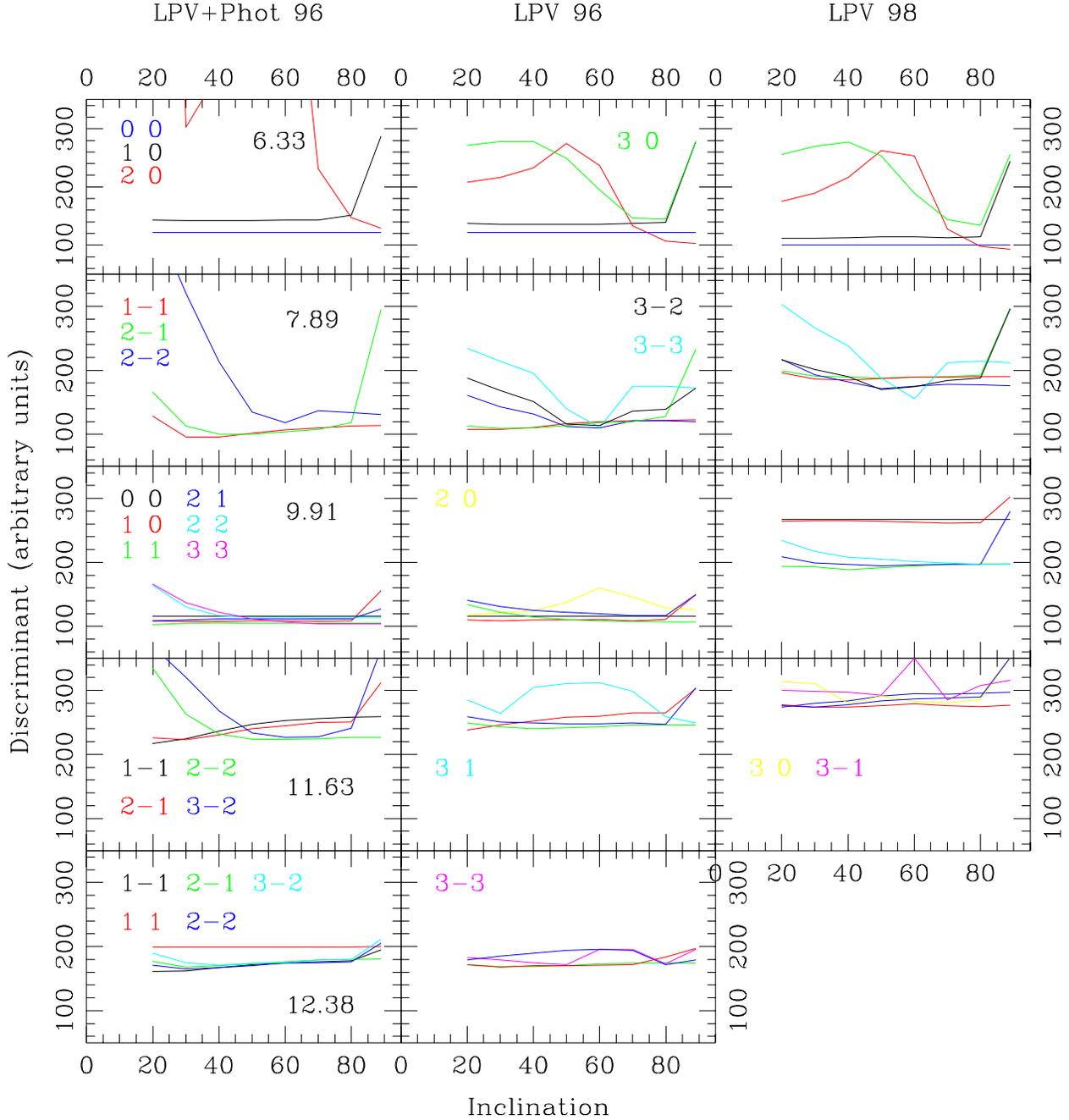}}
\caption[ ]{Behaviours of the discriminant of the best fitting modes versus
the inclination for the terms detected both spectroscopically and
photometrically. Left panels: discriminants computed by simultaneously
fitting line profile and light variations in the 1996 season. Central panels:
discriminant computed by fitting line profile variations in the 1996 data only.
Right panels: discriminants computed by fitting line profile variations
in the 1998 data. The panels of the same line concern the same term whose frequency
is reported in the leftmost panel.}
\label{colo}
\end{figure*}

The identification of the
modes which were independently detected both by photometry and spectroscopy
was performed looking for all the possible $\ell,m$ combinations up to $\ell=4$
and with inclinations of the rotation axis between 20 and 90 degrees
(the lower limit was fixed considering that the star reaches the break--up
velocity at about $i=17^o$).
The main results are shown in Fig.~\ref{colo} where the horizontal panels
belong to the same mode whose frequency is reported in the leftmost panel.
In each panel the behaviours of the discriminants versus the inclinations
of the best fitting modes are reported. Each mode is identified by a different
colour.
Since in the 1998 season we have no photometric data, in order to allow a
comparison of the results the 1996 data were analyzed fitting
simultaneously line profile and light variations (left panels)
and only line profile variations (central panels).
The panels on the right contain the discriminant computed from the 1998
spectroscopic data. In the first case the models allowed
for both velocity and flux
variations, while in the second case only velocity variations were considered.
In the 1998 data the 12.38 \cds term was not detected and therefore the
corresponding panel is missing.
The simultaneous fitting of velocity and flux variations made  
the results shown in the leftmost panels more reliable.
Those in the other two
panels are useful only to check how reasonable it is to neglect flux
variations to model line profile variations (and hence if we can avoid getting
simultaneous light and spectroscopic data) and
the consistency between the two seasons' data.
The general conclusion is that the agreement between the best fitting modes
in the two seasons' data is reasonable because
the best fitting modes are the same, with a few exceptions. However we can see
that by fitting
line profile variations only we tend to assign a higher degree to the detected
modes. The light curves play a decisive role in the
mode typing since the photometric amplitude is very sensitive to the
cancellation effects originated by the high--degree modes.    
 For instance for the 6.33 \cds term the possibility of
identifying it as $3,0$
is completely ruled out if we also take into account light variations.

As regards possible identification, we can give reasonable
estimates for two terms only. The terms at 9.91, 11.63, and 12.38 \cds are 
the more uncertain since many modes supply comparable results.
The term at 6.33~\cds is certaintly axissymmetric (all the proposed
identifications have $m$=0) and its most probable
identification is $\ell=0$, $m=0$, unless the star is seen almost exactly
equator--on. In this case $\ell=2$, $m=0$ could also be possible.
The term at 7.89~\cds has $\ell=1$, $m=-1$ or $\ell=2$, $m=-1$ as the most
probable identifications, although we cannot completely rule out the possibility
$\ell=2$, $m=-2$.

Because of the mode involved, the present discriminants  do not help
very much to decide on the most probable inclination of the rotational axis,
which however will be constrained on the basis of different
considerations later.

\subsection {High degree prograde modes}
\begin{figure}
\resizebox{\hsize}{!}{\includegraphics{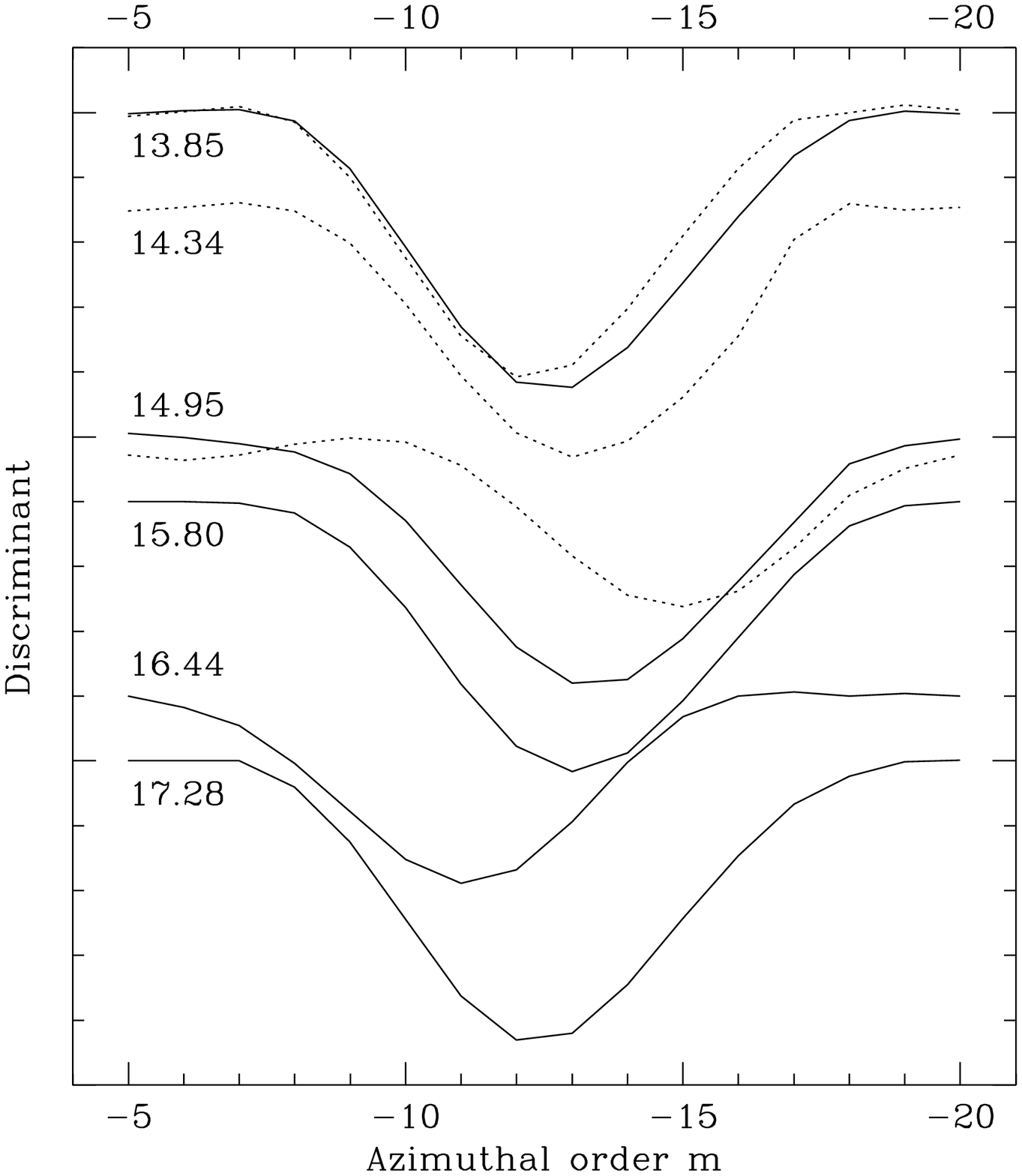}}
\caption[ ]{Discriminants of the six high degree prograde modes as a function
of the azimuthal order $m$ computed from the 1996 (dashed lines) and 1998
(solid lines) data. These values are relative to $i=55^o$ and $\ell=-m$.
But the discriminants are rather insensitive to $i$ and are comparable for any
$\ell\geq\mid m\mid$.}
\label{high}
\end{figure}

Six of the detected terms belong to this group: 13.85 and 14.95~\cds (which
were detected both in the 1996 and in the 1998 data), 14.34~\cds (which was detected in
the 1996 data) and 15.80, 16.44 and 17.28~\cds (which were detected in the 1998 data).
In order to fit the line profile variations produced by these terms all the
prograde modes with $5\leq \ell \leq 20$ were explored,
either taking into account flux variations or not.
These high--$\ell$~modes are not very sensitive to flux variations.
In turn, line profile variations are
rather insensitive to the inclination and to the $\ell$ degree, while they
were strongly dependent upon the azimuthal order. Figure~\ref{high}
shows the discriminant of these six terms as a function of $m$ and computed
for $i=55^o$ and $\ell=\mid m\mid$.  However as was said above the inclination is
rather unimportant, as for a fixed mode the discriminant is flat at minimum
and increases appreciably only for $i \leq 35^o$.
In Fig.~\ref{high} the solid lines indicate the discriminants derived from the
1998 data
while the dashed lines indicate those obtained from the 1996 data, and they are in
agreement.
From Fig.~\ref{high} we can also estimate the $m$ values reported in Tab.~\ref{summa}.

\begin{figure}
\resizebox{\hsize}{!}{\includegraphics{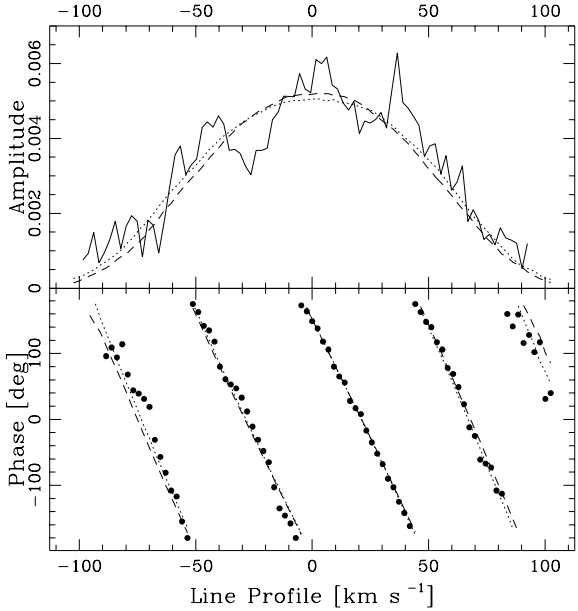}}
\caption[ ]{Behaviour of amplitude (solid line upper panel) and phase (dots,
lower panel) of the 17.28 \cds term across the line profile and their best
fits with $\ell=-m=12$ (dashed lines) and $\ell=-m=13$ (dotted lines).
}
\label{part}
\end{figure}

As an example of how the fits of the models to the observational
quantities look we show in Fig.~\ref{part} the behaviour of the amplitude and phase
across the line profile of the 17.28 \cds mode with the best fitting solutions
for $\ell=12, m=-12$ (dashed lines) and $\ell=13, m=-13$ (dotted lines).

\begin{table*}
\caption[]{Detection (Y: positive; N: negative) and typings of the pulsation
 modes observed
in the photometric (P, 1980 and 1996) and spectroscopic (S, 1996 and
1998) datasets of BV Cir. No mode identification is possible for the modes
detected only photometrically (lower panel).}
\begin{flushleft}
\begin{tabular}{rllccc}
\hline\noalign{\smallskip}
Frequency & Technique & \multicolumn{1}{c}{Typing} & & \multicolumn{2}{c}{Detection}\\
\cline{5-6}
[\cd]&
\multicolumn{2}{c}{ } & & 1996 & 1998 \\
\hline  
\noalign{\smallskip}
1.48  & S     & $m$=7 or 8                        & & Y & Y \\
6.33  & S~P   & $\ell$=0, $m$=0                   & & Y & Y \\
7.89  & S~P   & $\ell$=1 or 2, $m$= --1 or --2    & & Y & Y \\
9.91  & S~P   & $0\leq\ell\leq3, m>0$             & & Y & Y \\
11.63 & S~P   & $\ell=2\pm1$, $m$=--1 or --2      & & Y & Y \\
12.38 & S~P   & $-1\leq\ell\leq3, m<1, m\neq0$    & & Y & N \\
13.85 & S     & $m$= --12 or --13                 & & Y & Y \\
14.34 & S     & $m$= --13                         & & Y & Y \\
14.95 & S     & $m$= --14                         & & Y & Y \\
15.80 & S     & $m$= --13                         & & N & Y \\
16.44 & S     & $m$= --12                         & & N & Y \\
17.28 & S     & $m$= --12 or --13                 & & N & Y \\
\hline
\noalign{\smallskip}
\multicolumn{3}{c}{ } & & 1980 & 1996\\
\cline{5-6}
10.23 & P     &                                   & & Y & Y \\
12.29 & P     &                                   & & Y & Y \\
11.077& P     &                                   & & Y & N \\
11.128& P     &                                   & & Y & N \\
\hline
\end{tabular}
\end{flushleft}
\label{summa}
\end{table*}
\section{Discussion and Conclusions}
BV Circini has a lot of excited pulsation modes with observed frequencies
ranging from low to high values. Nine frequencies were detected from
photometric data (1980 and 1996) and 13 from spectroscopic (1996 and 1998)
ones, 5 of them are in common
in both datasets. The amplitudes of some modes are dramatically changing
and this is particularly evident from the two spectroscopic datasets, in
particular the strongest spectroscopic mode in the 1998 data at 17.28 \cds,
was not detectable in the previous dataset.

Six of the detected modes are high azimuthal order prograde modes with
$-14\leq m \leq -12$ and  one is retrograde with $m=$7 or 8.
For only two of the modes detected by simultaneous photometry and spectroscopy
it has been possible to suggest a complete $(\ell, m)$ 
identification, while 
better $S/N$ spectroscopic data are needed for the others. The two identified modes
are the one at 6.33 \cds (the photometric dominant mode),
which can be  identified as the  $\ell=0$ (radial mode).
The other is the one at 7.89 \cds which is a low degree prograde mode
with $\ell=1$ or $2$ and $m=-1$. The proposed typings of the pulsation
modes of BV Cir are summarized in Tab.~\ref{summa}.

By using the physical parameters computed in Sect.~2 we can evaluate the
pulsation constant of the 6.33~\cds mode, i.e. $Q=0.032\pm0.005$~d.
Therefore, this mode is probably the radial fundamental in agreement
with the suggestion by Kurtz (1981). If we take this result for granted,
we can use this information to narrow the uncertainties on some physical
parameters. In fact, assuming for this mode the
theoretical value of $Q=0.033\pm 0.001$~d as pulsation constant, we get
from its definition 
$\overline{\rho}/\overline{\rho}_{\sun}=0.044\pm 0.003$ and also
$\log g=3.63\pm0.04$. Moreover if we combine the mean density with the
radius estimate we get $M=2.04\pm 0.12 M_{\sun}$, again in excellent
agreement with the value derived from the evolutionary models (see Sect.~2).
\begin{table}
\caption[]{Basic physical parameters of BV Cir}
\begin{flushleft}
\begin{tabular}{ll}
\hline\noalign{\smallskip}
Parameter & Estimated value \\
\hline
Mass & $2.04\pm 0.12 M_{\sun}$\\
Effective temperature & $7260\pm100^oK$\\
Absolute bolometric magnitude & $0.97\pm 0.17$\\
$\log g$& $3.63\pm0.04$\\
Radius &$3.59\pm0.30 R_{\sun}$\\
Equatorial rotational velocity & $111\pm 12$ \kms\\
Rotational frequency & $0.61\pm0.07$ d$^{-1}$\\
\hline
\end{tabular}
\label{para}
\end{flushleft}
\end{table}

Hence by comparing the information from pulsation properties
(as derived from our analysis), {\sc hipparcos}
parallax and effective temperature (as derived from multicolour photometry),
 we determined  physical parameters which are
perfectly consistent (Tab.~\ref{para}.)

We have remarked in the previous section that the discriminants of
the detected modes are not particularly sensitive to the
inclination of the rotation axis.
However we can constrain this value using the azimuthal orders
derived for the high degree modes and the fact that the 6.33 \cd
mode is the radial fundamental, because in this case their
frequencies in the stellar reference frame should be larger than
the radial fundamental one. Among the prograde modes the critical
one is the 13.85~\cds, which with $m=-12.5\pm 1$ imposes $i\geq
62^o\pm10^o$, while the 1.48~\cds with $m=7.5\pm1$ gives $i\leq
56^o\pm 12^o$. By combining the two results we obtain $i\sim
60^o$. Consequently the equatorial rotation velocity is about 111
kms$^{-1}$ and the rotation frequency $\Omega$ about 0.61~rev~d$^{-1}$.

Let us finally deal with a few considerations about the technique to fit line profile
variations. We have seen that in order to get an unambiguous discrimination
between the possible identifications we need data with a rather high $S/N$.
The present data, with a $S/N$ of the order of 250, supply useful
information for the strongest spectroscopic terms only. Since the
observed amplitudes are typical of $\delta$ Scuti stars, it is not
unreasonable to believe that, in order to do a good job, data with a
$S/N$ of 500 or better could be necessary.

We have also explored the difference between fitting line profile variations
considering velocity fields only or taking also into account flux variations.
We have seen that in the first case we tend to overweight the higher degree
modes when we try to identify low degree modes.
For these modes it is therefore important to consider flux variations too,
and in this case, in order to properly constrain the fitted model,
it is necessary to have simultaneous light variations.

\acknowledgements{The authors wish to thank D.~Kurtz for kindly putting
his data at their disposal and  Jacques Vialle for checking the English
form of the manuscript.}

\end{document}